# Correlation between the tolerance factor and phase transition in $Ln_{4-x}Ln'_xNi_3O_{10}$ (*Ln and Ln'* = La, Pr and Nd; *x* = 0, 1, 2 and 3)

Shangxiong Huangfu, Xiaofu Zhang, Andreas Schilling

Department of Physics, University of Zurich, Winterthurerstrasse 190 CH-8057 Zurich Switzerland

*Abstract*

We have synthesized a series of the Ruddlesden–Popper nickelate solid solution $Ln_{4-x}Ln'_xNi_3O_{10}$ (*Ln and Ln'* = La, Pr and Nd; *x* = 0, 1, 2 and 3) via the citrate precursor method in different reacting atmospheres. Both the electronic-transport and magnetization measurements on these samples show well-defined phase transitions at temperatures between 135 K and 165 K. These transition temperatures as well as the room-temperature resistivities strongly correlate with the Goldschmidt tolerance factor *t*, irrespective of the combination of the magnetic rare-earth ions with non-magnetic $La^{3+}$. We conclude that these changes of the electronic properties are mostly a consequence of the distortion of the $NiO_6$ octahedra at the phase transition which is intimately related to the tolerance factor *t*, but appear to be rather insensitive to the average magnetic moment of the rare-earth ions upon exchanging La by Pr or Nd.

**Introduction**

Due to the strongly correlated *d*-electrons, some layered transition-metal oxides give rise to various phases with different intriguing electronic and magnetic properties [1-4]. The transitions between these phases, e.g. metal-insulator transitions [5,6], paramagnetic-ferromagnetic transitions [7], or Mott-insulator-superconductor transitions [8-10], are driven by various parameters such as temperature, pressure, electric current, magnetic field, chemical composition, etc. [5,8,11-14]. Generally, these phase transitions are associated with distinct changes of the physical properties, such as heat capacity, resistivity, or magnetization. The Ruddlesden–Popper (R-P) nickelates $Ln_{n+1}Ni_nO_{3n+1}$ are constituted by perovskite-like structures piled by a number ($n$) of infinite $NiO_6$-octahedron-based layers, allowing for many different species at the *Ln* site (rare-earth-metals, La, Pr, and Nd, and alkaline-metals such as Sr and Ba). Therefore, one may expect that there are many variants of distorted $NiO_6$ octahedra and tunable mixed-valence states of nickel with values between +2 and +3 in this series of compounds. Due to the diverse *d*-orbital states of the nickel ions, which in turn may lead to a variety of interesting physical phenomena, these R-P nickelates have recently gained intensive attention.

By doping or substituting, the R-P rare-earth nickelates show abundant and tunable physical properties and phase changes. The $La^{3+}$ ions in $La_2NiO_4$ ($n = 1$), for instance, can be partially replaced by $Ba^{2+}$ ions or $Sr^{2+}$ ions. In the resulting solid solution of $La_{2-x}Sr_xNiO_4$, a maximum of the tetragonality ratio $c/a$ is observed at $x = 0.5$ [15], and the semiconductor-metal crossover temperature monotonically decreases with increasing $x$ from 0 to 1.2 [16], while $LaBaNiO_4$ remains insulating down to the zero temperature, instead [17]. Moreover, the solid solution of $LaSr_{1-x}Ba_xNiO_4$ shows electronic transport properties ranging from a metallic state ($x = 0$) to an insulating state ($x = 1$) [18]. Another example, the perovskite $LnNiO_3$ ($Ln$ = Y, Pr to Lu, $n = \infty$), features metal-insulator phase transitions [19-21]. In the solid solution $Nd_{1-x}Ln_xNiO_3$ ($Ln$ = Sm, Eu, Gd, Er, and Yb), the metal-insulator transition temperatures can be varied over a large temperature range from 200 K to 450 K by substitution of the rare-earth ions [22].

The $Ln_4Ni_3O_{10}$ ($Ln$ = La, Pr, Nd, $n$ = 3), which has recently attracted a lot of interest due its structural similarity to some high-$T_c$ superconductors, shows phase transitions that are accompanied by a resistivity shift at transition temperatures between 128 K and 165 K [23-25]. The resistivity of $La_4Ni_3O_{10}$, exhibits an anomaly in both powder and single crystal samples at $T_{pt}$ ~ 128 K, indicating a possible metal-metal phase transition [24,26], which is even more pronounced in $Pr_4Ni_3O_{10}$ and $Nd_4Ni_3O_{10}$ [23,25]. In single crystals of $Pr_4Ni_3O_{10}$, even a metal-insulator-like transition is observed along the $c$-axis at $T_{pt}$ ~ 157 K [27]. $La_4Ni_3O_{10}$, $Pr_4Ni_3O_{10}$ and $Nd_4Ni_3O_{10}$ all show certain changes in their crystal structure at the respective transition temperatures [25,27,28], and different magnetization behaviours between the low-temperature and the high-temperature phases [23,25,27,29].

In this work, we have synthesized solid solution samples of $Ln_{4-x}Ln'_xNi_3O_{10}$ (*Ln and Ln'* = La, Pr and Nd) with $x$ = 0, 1, 2, and 3. The successful substitution of La, Pr and Nd was confirmed by powder X-ray diffraction (XRD) and energy-dispersive X-ray spectroscopy (EDX). We have performed systematic measurements on their electronic and magnetic properties. These measurements revealed well-defined phase transition temperatures associated with changes of the physical properties for all values of *x*, some of which are found to be strongly correlated with the Goldschmidt tolerance factor *t*.

**Experimental method**

The $Ln_{4-x}Ln'_xNi_3O_{10}$ samples (*Ln and Ln'* = La, Pr and Nd; $x$ = 0, 1, 2, and 3) were synthesized via the citrate precursor method [27,30], the details of which are given in the Supplemental Material [31-34]. These precursors have been annealed at 1100 ℃ in oxygen atmosphere for 24 hours, except for $La_4Ni_3O_{10}$ (which does not require oxygen atmosphere) and $Nd_4Ni_3O_{10}$ (which needs a reacting time of 72 hours). The compacted and pelleted samples for electronic-transport measurements were obtained by pressing the powder and annealing again at 1100 °C in oxygen atmosphere for 24 hours. All samples were characterized by powder XRD at room temperature (Fig. 1a), and the results show sharp peaks without visible impurities. The elemental compositions of all samples were confirmed by EDX (see Figs S2 to S13 in the Supplemental Material) [31]. According to our thermo-gravimetric analysis, we can confirm an almost stoichiometric oxygen content for all compounds with values around 10.0 within 1% or better (see Table SI in the Supplemental Material) [31].

The resistivities of the $Ln_{4-x}Ln'_{x}Ni_3O_{10}$ samples were measured with a Physical Property Measurement System (PPMS, *Quantum Design Inc.*), and a standard four-probe technique was employed in the temperature range from 10 K to 300 K. The magnetic properties were studied by using a Magnetic Properties Measurement System (MPMS 3, *Quantum Design Inc.*) from 10 K to 300 K with an external magnetic field of 0.1 T.

**Results**

As a result of distortions of both shape and arrangement of the $NiO_6$ octahedra, the crystal structures of the $Ln_{4-x}Ln'_{x}Ni_3O_{10}$ solid solution show a relatively low symmetry. Based on available experimental data, the unit cell of $La_4Ni_3O_{10}$ can be equally well fitted to two types of space groups: *P2$_1$/a* (monoclinic) and *Bmab* (orthorhombic) [28], while the $Pr_4Ni_3O_{10}$ and $Nd_4Ni_3O_{10}$ can only be well described by the space group of *P2$_1$/a* (monoclinic) [28]. Figure 1(a) shows the powder XRD patterns for all samples involved in this study, and in Fig. 1(c), we summarized all the changes of the corresponding lattice parameters (see Supplemental Material for details) [31]. In the $La_{4-x}Pr_xNi_3O_{10}$ series, besides the wholistic shift of these patterns that is ascribed to the expansion of unit cells, there are certain specific peak changes as $x$ varies from 0 to 4. In Fig. 1(b), the (117) and (002) peaks ($2\theta \sim 32.5$ °) separate gradually as the La content increases, while the (228) and (1115) peaks ($2\theta \sim 54.7$ ° and $\sim 55.4$ °), on the contrary, gradually merge. A similar tendency is also found for the $Nd_{4-x}La_xNi_3O_{10}$ samples. In the $Pr_{4-x}Nd_xNi_3O_{10}$ series, however, there is no analogous peak change other than a systematic shift due to the lattice contraction as the Nd content increases. This result is in accordance with the reported space groups [28]. The structural difference between $La_4Ni_3O_{10}$ and $Pr(Nd)_4Ni_3O_{10}$ can be mainly ascribed to the size difference between $La^{3+}$, $Pr^{3+}$, and $Nd^{3+}$, with average radius of 1.21 Å, 1.17 Å, and 1.16 Å in eight and ten coordinations, respectively [35].

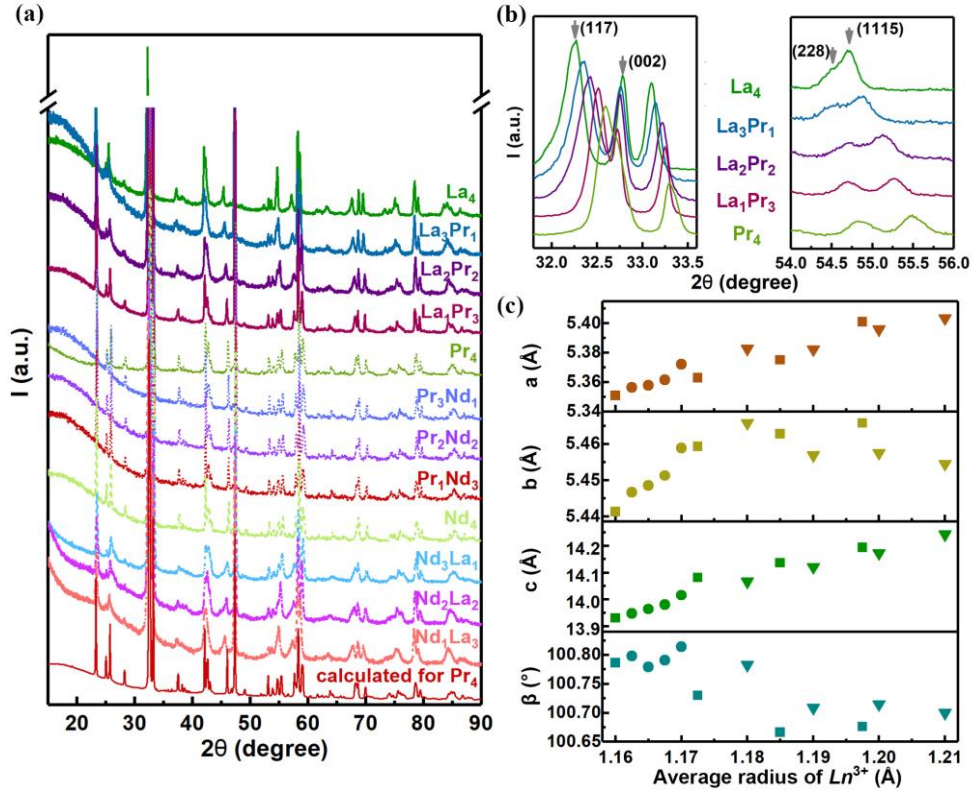

Fig. 1 (a) XRD patterns of the solid solution $Ln_{4-x}Ln'_xNi_3O_{10}$ samples and calculated pattern from the $Pr_4Ni_3O_{10}$ single crystal data [27]; solid curves represent $La_{4-x}Pr_xNi_3O_{10}$ ($x$ = 0, 1, 2, 3), dotted curves $Pr_{4-x}Nd_xNi_3O_{10}$ ($x$ = 0, 1, 2, 3), and dashed curves $Nd_{4-x}La_xNi_3O_{10}$ ($x$ = 0, 1, 2, 3); (b) Bragg peaks for $La_{4-x}Pr_xNi_3O_{10}$ samples at $2\theta \sim 31.6°$ to $\sim 33.8°$ (left) and at $2\theta \sim 54.0°$ to $\sim 56.0°$ (right), from top to bottom arranged from La to Pr; (c) Lattice parameters of the series $Ln_{4-x}Ln'_xNi_3O_{10}$ as functions of the average radius of $Ln^{3+}$, according to the monoclinic space group $P2_1/a$; the triangle, round, and square symbols represent $La_{4-x}Pr_xNi_3O_{10}$ ($x$ = 0, 1, 2, 3), $Pr_{4-x}Nd_xNi_3O_{10}$ ($x$ = 0, 1, 2, 3), and $Nd_{4-x}La_xNi_3O_{10}$ ($x$ = 0, 1, 2, 3), respectively.

To characterize the low-temperature phase transitions at temperatures between 135 K and 165 K, we performed electronic-transport measurements on the pelleted samples of the solid solution $Ln_{4-x}Ln'_xNi_3O_{10}$. Figure 2 shows the normalized resistivities $\rho(T)/\rho(300K)$ as a function of temperature from 10 K to 300 K. All samples show metallic behaviours in both the high-temperature and the low-temperature phases, which are consistent with the previous results for the corresponding parent compounds [23]. Besides, we also observe clear anomalies in the resistivity for all compositions at characteristic temperatures $T_{pt}$, indicating phase transitions with a systematically increasing $T_{pt}$ (from $\sim 135$ K for $La_4Ni_3O_{10}$ to $\sim 157$ K for $Pr_4Ni_3O_{10}$ and to $\sim 161$ K for $Nd_4Ni_3O_{10}$) when substituting the rare-earth ions $La^{3+}$ by $Pr^{3+}$, $Pr^{3+}$ by $Nd^{3+}$ and $La^{3+}$ by $Nd^{3+}$, respectively. The associated relative resistivity changes at $T_{pt}$, $\Delta\rho(T_{pt})/\rho(300K)$, show a trend to gradually increase from ~ 2 % ($La_4Ni_3O_{10}$) to ~ 9 % ($Pr_4Ni_3O_{10}$) and eventually to ~ 11% ($Nd_4Ni_3O_{10}$) with decreasing temperature. However, these numbers represent a polycrystalline average and may be affected by the connectivity of the individual grains, and they can therefore not represent the intrinsic anisotropic values obtained on single-crystals [27].

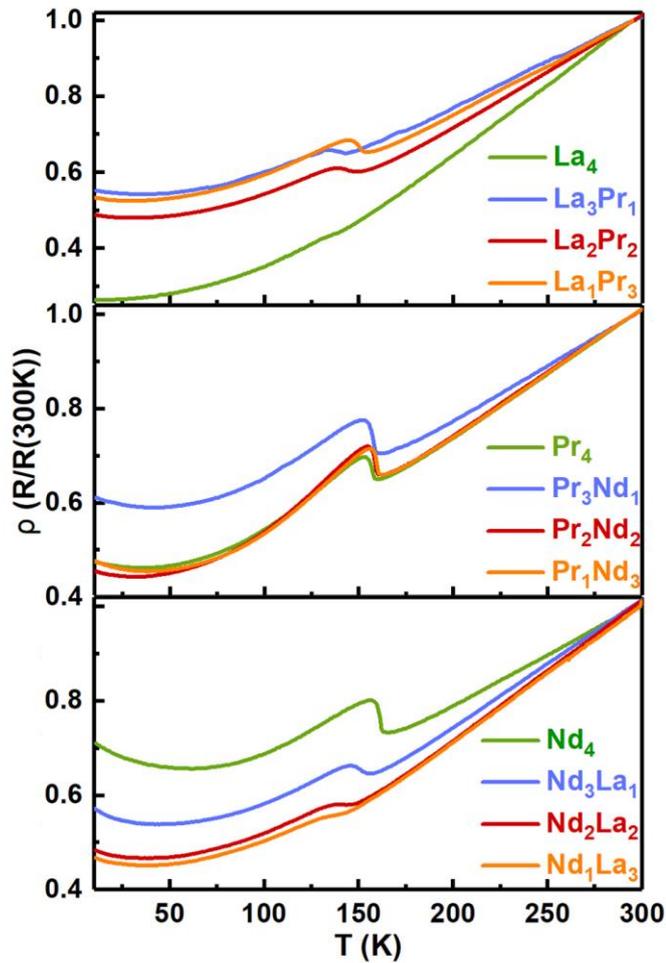

Fig. 2 Zero-field resistivities, normalized to the resistance at 300 K, of the solid solution $Ln_{4-x}Ln'_xNi_3O_{10}$ as functions of temperature; upper, middle, and bottom show $La_{4-x}Pr_xNi_3O_{10}$ ($x = 0, 1, 2, 3$), $Pr_{4-x}Nd_xNi_3O_{10}$ ($x = 0, 1, 2, 3$), and $Nd_{4-x}La_xNi_3O_{10}$ ($x = 0, 1, 2, 3$), respectively.

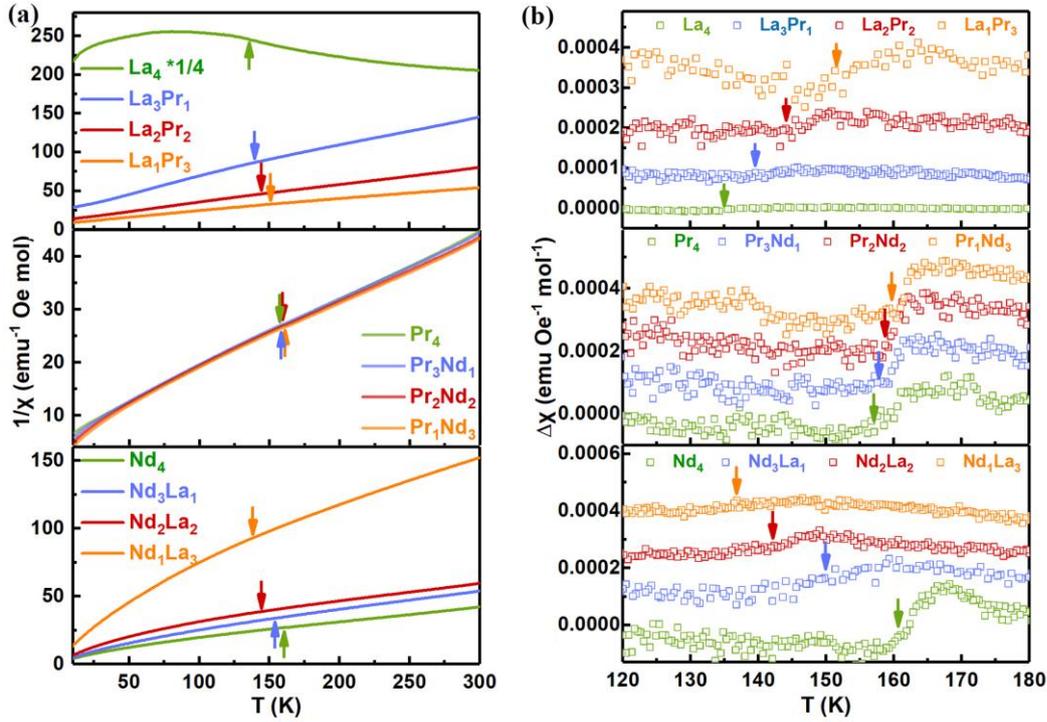

Fig. 3 (a) $1/\chi$ of the solid solution $Ln_{4-x}Ln'_x\mathrm{Ni}_3\mathrm{O}_{10}$ as functions of temperature; the data for $\mathrm{La}_4\mathrm{Ni}_3\mathrm{O}_{10}$ are multiplied by 1/4 for clarity; (b) The same $\chi(T)$ data after subtraction of a fitted Curie-Weiss background (3$^{rd}$ order polynomial background for $\mathrm{La}_4\mathrm{Ni}_3\mathrm{O}_{10}$) in a temperature window of +/- 50 K around the respective $T_{pt}$ (marked with arrows). The data are offset by 0.00012 emu Oe$^{-1}$ mol$^{-1}$ for clarity.

We also measured the magnetization for the solid solution $Ln_{4-x}Ln'_x\mathrm{Ni}_3\mathrm{O}_{10}$ at temperatures between 10 K and 300 K [Fig. 3(a)] to characterize the changes in the magnetic properties as functions of their $Ln$ constituents. The magnetic-susceptibility data of the zero-field-cooling (ZFC) and the field-cooling (FC) essentially overlap within the measured temperature range. The magnetic susceptibility of $\mathrm{La}_4\mathrm{Ni}_3\mathrm{O}_{10}$ decreases with the decreasing temperature in the high-temperature phase, while in the low-temperature phase, the susceptibility increases as the temperature decreases, which is similar to previous reports [23, 29]. None of the phases follows a single simple Curie-Weiss law over the whole temperature range. However, except for $\mathrm{La}_4\mathrm{Ni}_3\mathrm{O}_{10}$, the temperature dependence of the magnetic susceptibilities $\chi(T)$ of all the other compositions in the $Ln_{4-x}Ln'_x\mathrm{Ni}_3\mathrm{O}_{10}$ series can be well fitted by a Néel-type of Curie-Weiss law for the high-temperature phase, $\chi(T) = \frac{C}{T+\Theta} + \chi_0$, demonstrating a paramagnetic behaviour at these temperatures. The temperature window used for this fitting procedure was chosen between $T_{pt}$ +15 K and 300 K. The corresponding results are listed in Table I and shown in Fig. 4. If we take the magnetic-susceptibility data of La$_4$Ni$_3$O$_{10}$ as a measure for possible Ni containing paramagnetic impurities, we obtain in the worst case a contribution of $\approx$ 0.01 emu K Oe$^{-1}$ mol$^{-1}$ to the listed values of $C$ (see Supplementary Material). Apart from the changes in $\chi(T)$ at $T_{pt}$, we have no indications for further magnetic transitions in the whole investigated temperature range.

Table I

| | $C$ (emu K Oe$^{-1}$ mol$^{-1}$) | $\chi_0$ (emu Oe$^{-1}$ mol$^{-1}$) | $\Theta$ (K) | $t$ | $T_{pt}$ (K) |
|---|---|---|---|---|---|
| La$_4$Ni$_3$O$_{10}$[#] | - | 2.4(1) × 10$^{-3}$ | - | 0.9140 | 135.5 |
| La$_3$Pr$_1$Ni$_3$O$_{10}$ | 1.88(3) | 1.6(1) × 10$^{-3}$ | 50(2) | 0.9107 | 139.5 |
| La$_2$Pr$_2$Ni$_3$O$_{10}$ | 4.00(9) | 1.3(3) × 10$^{-3}$ | 50(3) | 0.9071 | 144.0 |
| La$_1$Pr$_3$Ni$_3$O$_{10}$ | 5.75(2) | 2.1(4) × 10$^{-3}$ | 51(4) | 0.9036 | 151.0 |
| Pr$_4$Ni$_3$O$_{10}$ | 7.6(3) | 7(4) × 10$^{-4}$ | 51(4) | 0.9000 | 157.0 |
| Pr$_3$Nd$_1$Ni$_3$O$_{10}$ | 7.6(3) | 1.5(4) × 10$^{-3}$ | 55(4) | 0.8991 | 157.8 |
| Pr$_2$Nd$_2$Ni$_3$O$_{10}$ | 7.6(2) | 1.7(4) × 10$^{-3}$ | 55(3) | 0.8982 | 158.8 |
| Pr$_1$Nd$_3$Ni$_3$O$_{10}$ | 7.6(3) | 1.8(5) × 10$^{-3}$ | 52(4) | 0.8973 | 159.6 |
| Nd$_4$Ni$_3$O$_{10}$ | 7.7(3) | 2.9(6) × 10$^{-3}$ | 61(5) | 0.8964 | 160.5 |
| La$_1$Nd$_3$Ni$_3$O$_{10}$ | 5.2(2) | 3.6(4) × 10$^{-3}$ | 46(4) | 09009 | 150.0 |
| La$_2$Nd$_2$Ni$_3$O$_{10}$ | 3.5(2) | 3.2(4) × 10$^{-3}$ | 44(7) | 0.9053 | 142.0 |
| La$_3$Nd$_1$Ni$_3$O$_{10}$ | 1.74(4) | 1.7(1) × 10$^{-3}$ | 54(3) | 0.9098 | 136.5 |

[#] The Pauli-paramagnetic susceptibilities of La$_4$Ni$_3$O$_{10}$ are estimated directly from the measured data right below 300 K.

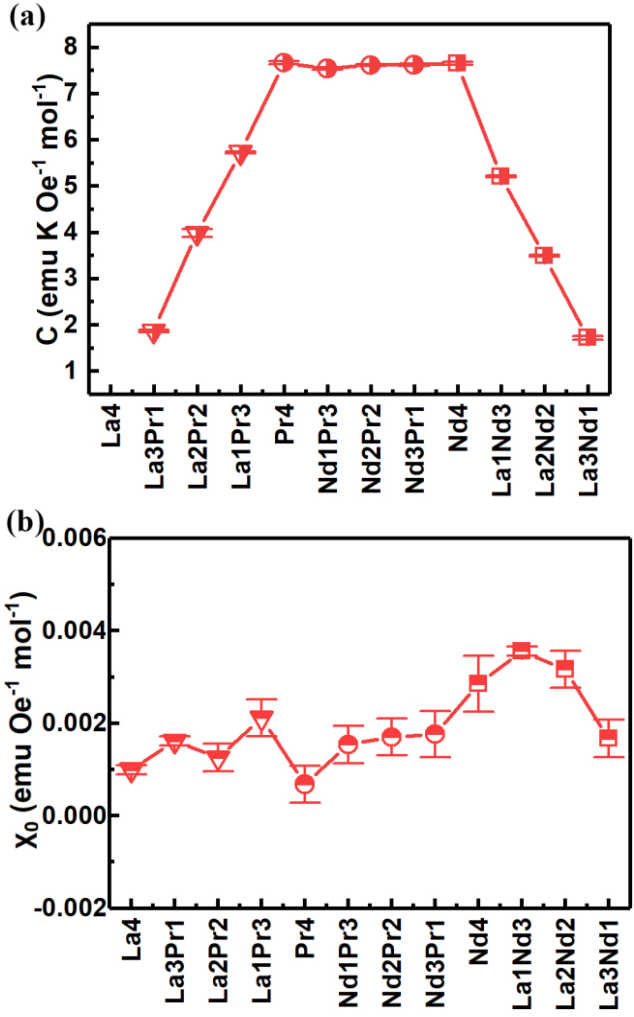

Fig. 4 Curie-constants $C$ (a) and Pauli-paramagnetic susceptibilities $\chi_0$ (b) as functions of element contents, calculated from the magnetic susceptibilities of the high-temperature phase; the triangle, round, and square symbols represent $La_{4-x}Pr_xNi_3O_{10}$ ($x$ = 0, 1, 2, 3), $Pr_{4-x}Nd_xNi_3O_{10}$ ($x$ = 0, 1, 2, 3), and $Nd_{4-x}La_xNi_3O_{10}$ ($x$ = 0, 1, 2, 3), respectively.

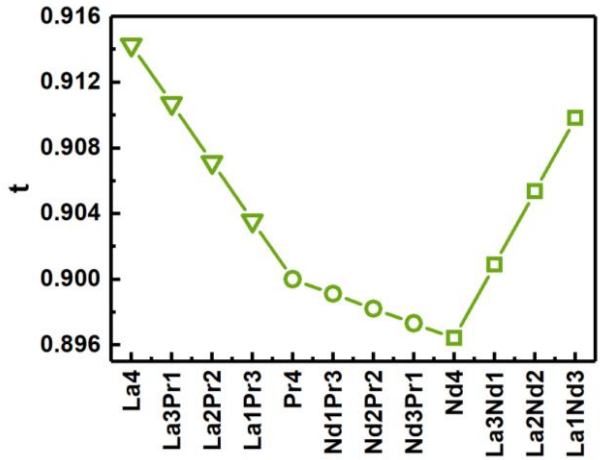

Fig. 5 Tolerance factor $t$ of the solid solution $Ln_{4-x}Ln'_xNi_3O_{10}$ as a function of element content; the triangle, round, and square symbols represent $La_{4-x}Pr_xNi_3O_{10}$ ($x$ = 0, 1, 2, 3), $Pr_{4-x}Nd_xNi_3O_{10}$ ($x$ = 0, 1, 2, 3), and $Nd_{4-x}La_xNi_3O_{10}$ ($x$ = 0, 1, 2, 3), respectively.

**Discussion**

The Goldschmidt tolerance factor $t$ is widely used to estimate the stability of the rare-earth perovskite-like structures, with $t = (r_{rare-earth} + r_O)/\sqrt{2}(r_{Ni} + r_O)$ [36], where $r_{rare-earth}$, $r_{Ni}$, and $r_O$ represent the average ionic radius of the rare-earth, nickel, and oxygen ions in the corresponding coordinations, respectively. It is found that, the values of $t$ in R-P type of compounds range from 0.86 to 0.99 [20,37,38]. By linear interpolation of the reported ionic radius [35, 39], the $t$ for all the samples of the solid solution $Ln_{4-x}Ln'_xNi_3O_{10}$ can be estimated. Upon changing $Ln^{3+}$, from $La^{3+}$ with the largest ionic radius, to $Nd^{3+}$ with the smallest ionic radius, the tolerance factor $t$ decreases from 0.914 to 0.896 [Fig. 5]. Considering our experience in the syntheses of these compounds, it becomes more and more difficult to obtain $Ln_4Ni_3O_{10}$ as $Ln^{3+}$ changes from $La^{3+}$ to $Pr^{3+}$ ($Nd^{3+}$), and from $Pr^{3+}$ to $Nd^{3+}$, suggesting that a larger value of $t$ results in a more stable phase. The $t$ difference between $La_4Ni_3O_{10}$ (0.914) and $Pr_4Ni_3O_{10}$ (0.900) is larger than that between $Pr_4Ni_3O_{10}$ (0.900) and $Nd_4Ni_3O_{10}$ (0.896). In any case, we can expect that the structures of these compounds gradually change with $t$, which in turn should lead to a certain $t$ dependence of the physical properties in connection with the phase transition.

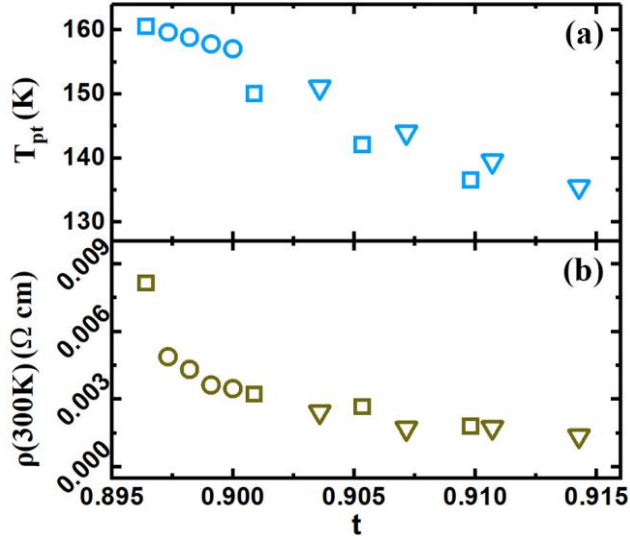

Fig. 6 Tolerance factor dependence of the phase transition temperatures $T_{pt}$ (a), the estimated absolute value of the resistivity at 300K (b); the triangle, round, and square symbols represent La$_{4-x}$Pr$_x$Ni$_3$O$_{10}$ ($x$ = 0, 1, 2, 3), Pr$_{4-x}$Nd$_x$Ni$_3$O$_{10}$ ($x$ = 0, 1, 2, 3), and Nd$_{4-x}$La$_x$Ni$_3$O$_{10}$ ($x$ = 0, 1, 2, 3), respectively; the error in $\rho$(300K) due to the contact geometry is ~ 30%.

Figure 6 shows the corresponding transition temperatures $T_{pt}$ and the absolute resistivities $\rho$(300 K) as functions of $t$. The $T_{pt}$ values were obtained from the peaks in $d\rho/dT$, as $\rho(T)$ discontinuously changes at the phase transition (see Fig. 2) [25, 27]. Except for La$_4$Ni$_3$O$_{10}$ without obvious shift of $\chi(T)$ at $T_{pt}$, corresponding $T_{pt}$ data taken from $\chi(T)$ are consistent with the $\rho(T)$ results within $\pm$ 0.5 K. As it is shown in Fig. 6(a), $T_{pt}$ is found to almost linearly decrease with $t$, with a slope of $dT_{pt}/dt \sim 1.6 \times 10^3$ K. Although the $t$ dependence of transition temperature $T_{pt}$ is similar to $Ln$NiO$_3$, the slope $dT_{pt}/dt$ of $Ln_4$Ni$_3$O$_{10}$ is approximately ten times smaller than that in $Ln$NiO$_3$ [20]. Similarly, the absolute values of the resistivity at room temperature [Fig. 6(b)] systematically decrease with $t$. Generally, for $t$ = 1, the distortion of the NiO$_6$ octahedra is known to be minimum [37], the Ni-O-Ni angles are close to 180º, and the overlap between the Ni $d$-orbitals and the O $p$-orbitals is maximized. This may explain the increasing metallicity in $Ln_{4-x}Ln'_x$Ni$_3$O$_{10}$ with $t$ in a similar way as it has been reported to be the case for $Ln$NiO$_3$ [20].

For most compositions, there is a step-like decrease $\Delta\chi$ of $\chi(T)$ at $T_{pt}$ with decreasing temperature, as we illustrate it in Fig. 3(b). The same behaviour has also been observed in Nd$_4$Ni$_3$O$_{10}$ powders [25], and we had interpreted this feature for Pr$_4$Ni$_3$O$_{10}$ single crystals in terms of a first-order phase transition [27]. As the entropy change $\Delta S$, the $\Delta \chi$ and the slope of the field-dependent $T_{pt}(B)$ are not independent from each other at such transitions, the sign of $\Delta \chi$ dictates the sign of $dT_{pt}/dB$ and *vice versa* [27], indicating a slight suppression of $T_{pt}$ with magnetic field in all cases, although its expected magnitude ($\approx 50$ mK in $B = 7$ T for Pr$_4$Ni$_3$O$_{10}$ [27]) is beyond the temperature resolution of our present data.

The Pauli-paramagnetic susceptibilities obtained from our Curie-Weiss type fits [Fig. 4(b)] do not show any clear trend and scatter around $\chi_0 \sim 2 \times 10^{-3}$ emu Oe$^{-1}$ mol$^{-1}$, but are an order of magnitude larger than the free-electron value $\chi_0^{f.e} \sim 10^{-4}$ emu Oe$^{-1}$ mol$^{-1}$ [27]. The corresponding Curie-constants $C$ show, as expected, large variations depending on the amount and the species of the rare-earth ions [Fig. 4(a)]. However, we observe systematically larger $C$ values than expected from a free-ion picture, namely $\approx 7.6$-$7.7$ emu K Oe$^{-1}$ mol$^{-1}$ for the end members Pr$_4$Ni$_3$O$_{10}$ and Nd$_4$Ni$_3$O$_{10}$, corresponding to $\approx 3.9$ $\mu_B$ per rare-earth ion, which has to be compared to the expected values $\approx 3.58$ $\mu_B$ for Pr$_4$Ni$_3$O$_{10}$ and $\approx 3.62$ $\mu_B$ for Nd$_4$Ni$_3$O$_{10}$, respectively. As our nonmagnetic La$_4$Ni$_3$O$_{10}$ samples prepared by similar methods do not show any sign of localized magnetic moments in the magnetic susceptibility (contributing to less than 0.01 emu K Oe$^{-1}$ mol$^{-1}$, see Supplementary Materials), and a similarly large Curie constant was found in single-crystalline Pr$_4$Ni$_3$O$_{10}$ prepared by different methods [27], we suggest that these values may be intrinsic and are not due to Ni-containing impurities. Larger-than-expected measured magnetic moments in rare-earth oxides have been reported for Nd$_2$WO$_6$ ($\approx 4.36$ $\mu_B$) [40], but also (and more important in the present context) for the *n* = 1 members of the R-P-nickelates Pr$_2$NiO$_4$ and Nd$_2$NiO$_4$, with magnetic moments up to $\approx 4.0$ $\mu_B$ for Pr$_2$NiO$_4$ [41] and $\approx 4.8$ $\mu_B$ for Nd$_2$NiO$_4$ [42], which had been attributed to interaction effects between the rare-earth and the Ni ions. On the other hand, the only systematic report on Nd$_4$Ni$_3$O$_{10}$ that we are aware of (with $C \approx 6.4$ emu K Oe$^{-1}$ mol$^{-1}$ [25]) does not at all seem to support our observation. However, we would like to mention that the oxygen contents of our Nd$_4$Ni$_3$O$_{10}$ samples ($\approx 9.93$) are closer to the ideal value 10.00 (see Table SI in the Supplemental Material) [31]) and differ from those reported in Ref. [25] ($\approx$ 9.85) for samples prepared by a different synthesis technique, which may have implications on the magnetic behaviour. Within this line of arguments, we also expect that the number of delocalized electrons due to the mixed valent $Ni^{2+/3+}$ is very sensitive to the exact oxygen content. As the number of such charge carriers is closely related to the Pauli-spin susceptibility via the electron-density of states, it is plausible that our value for $\chi$ in $Nd_4Ni_3O_{10}$ ($\approx 2.9 \times 10^{-3}$ emu $Oe^{-1}$ $mol^{-1}$) differs from that reported in Ref. [25] ($\approx 4.4 \times 10^{-3}$ emu $Oe^{-1}$ $mol^{-1}$).

Finally, we state that all the data given in Table I and shown in the figures were obtained on pressed powders, i.e., represent polycrystalline averages. Nevertheless, the quantitative and qualitative agreement of our data obtained for $Pr_4Ni_3O_{10}$ powders with those measured in $Pr_4Ni_3O_{10}$ single crystals [27] suggests that the trends described here reflect those of intrinsic properties of the whole series $Ln_{4-x}Ln'_xNi_3O_{10}$.

**Conclusion**

We have successfully synthesized a series of $Ln_{4-x}Ln'_xNi_3O_{10}$ (*Ln and Ln'* = La, Pr and Nd) compounds via the citrate precursor method in different reacting atmospheres. The phase transition temperatures $T_{pt}$ and the room-temperature resistivities $\rho(300K)$ vary systematically with the Goldschmidt tolerance factor *t*, and may even possibly be controlled by it. With increasing *t*, the compounds become more conducting, which we may attribute to a successively diminishing distortion of the $NiO_6$ octahedra which are responsible for the electron transport in these materials. Our results indicate that, while the associated changes of the Ni-O the bond lengths and Ni-O-Ni bond angles play a crucial role for the electronic structure of these compounds, the magnetism of the $Ln^{3+}$ ions upon replacing La by Pr or Nd is not a decisive factor.

**Acknowledgement**

This work was supported by the Swiss National Science Foundation under Grant No.20-175554.

*Supplemental Material for*

# Correlation between the tolerance factor and phase transition in $Ln_{4-x}Ln'_{x}Ni_3O_{10}$ (*Ln and Ln'* = La, Pr and Nd; *x* = 0, 1, 2 and 3)

Shangxiong Huangfu, Xiaofu Zhang, Andreas Schilling

Physics Department of the University Zurich, Winterthurerstrasse 190 CH-8057 Zurich Switzerland

## *Synthesis*

Powdered samples of $Ln_{4-x}Ln'_{x}Ni_3O_{10}$ (*Ln and Ln'* = La, Pr and Nd; *x* = 0, 1, 2, 3) were synthesized by a citric acid assisted sol-gel method. The reactants $La_2O_3$(99.9%; *Sigma-aldrich*), $Pr_6O_{11}$ (99.9%; *Sigma-Aldrich*), $Nd_2O_3$ (99.9%; *Fluka chemica*) and NiO (99.99%; *Sigma-Aldrich*) were weighted in stoichiometric ratios. Then, nitric acid (65% for analysis; *Emsure*) was used for dissolving these oxides, and clear green liquid were obtained. After adding citric acid monohydrate ($C_6H_8O_7*H_2O$; 99.5%; *Emsure*) in a molar ratio 1:1 with respect to the molar contains of cations, the resulting liquid were heated at around 300 ˚C on a heating plate, dried and decomposed into dark brown powders. Finally, ultrafine and homogeneous powder of $Ln_{4-x}Ln'_{x}Ni_3O_{10}$ were obtained by annealing the precursor powder at 1100 ˚C in flowing oxygen for 24 hours, except for $La_4Ni_3O_{10}$ (which does not require oxygen atmosphere) and $Nd_4Ni_3O_{10}$ (which needs a reacting time of 72 hours). The compacted pelleted samples were obtained by pressing the powder and annealing again at 1100 °C in oxygen atmosphere for another 24 hours.

The oxygen contents of the obtained compounds were determined by thermogravimetric reduction with 10% of $H_2/N_2$ gas, and the corresponding results are shown in Table SI.

## *Powder X-ray diffraction*

Powder X-Ray Diffraction (PXRD) data were collected at room temperature in transmission mode using a Stoe Stadi P diffractometer equipped with a $CuK_{\alpha 1}$ radiation (Ge(111) monochromator) and a DECTRIS MYTHEN 1K detector. The reflections of a main phase were indexed with a monoclinic cell in the space group $P2_1/a$ (No. 14). The Rietveld refinement analysis [32] of the diffraction patterns was performed with the package FULLPROF SUITE [33,34] (version March-2019). The structural model was taken from the single-crystal X-Ray diffraction refinement. Refined parameters were: scale factor, zero shift, transparency, lattice parameters, Pr and Ni atomic positions, and peak shapes as a Thompson–Cox–Hastings pseudo-Voigt function. A preferred orientation correction as a Modified March function was implemented in the analysis. The refinement data are listed in Table SI.

## *Energy-dispersive X-ray spectroscopy*

The elemental compositions are determined by means of energy-dispersive Xray (EDX) analysis integrated into the Zeiss Supra 50 VP scanning-electron microscope (SEM). The data taken on all the compacted samples of $Ln_{4-x}Ln'_xNi_3O_{10}$, confirmed the presence of La, Pr and Nd with elements contents that very in good agreement with the expected ratios (See Figs. S3 to S14).

*Transport measurements*

The resistivity measurements of the $Ln_{4-x}Ln'_xNi_3O_{10}$ pelleted samples were performed in a Physical Property Measurement System (PPMS, *Quantum Design Inc.*), and a standard four-probe technique was employed with 50μm silver wires attached with silver paint. All the measurements were performed with an applied current of $I = 0.5$ mA in the temperature range from 10 K to 300 K.

*Magnetization measurements*

The magnetic properties of $Ln_{4-x}Ln'_xNi_3O_{10}$ were studied with a Magnetic Properties Measurement System (MPMS 3, *Quantum Design Inc.*). In the temperature dependent magnetization ($M(T)$) measurements, both zero-field cooling and field cooling processes were measured in the temperature range of 10 K to 300 K with an external magnetic field of 0.1 T.

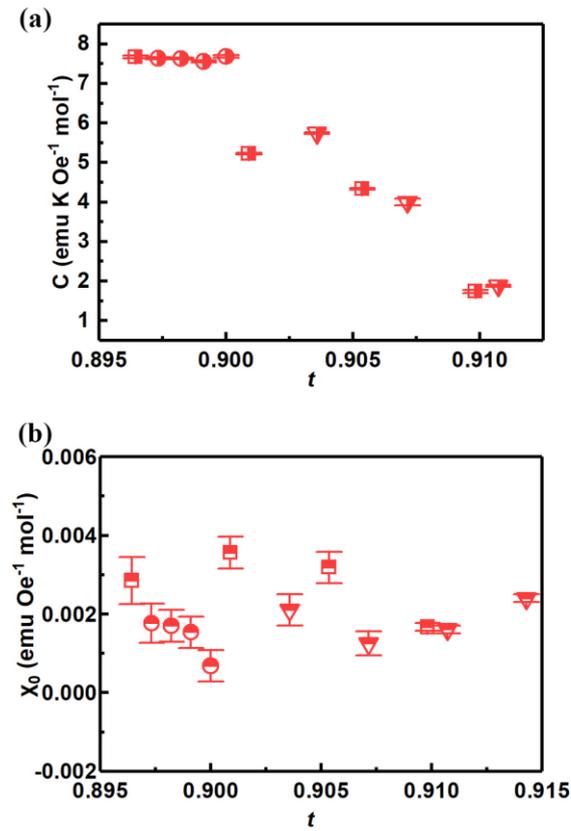

Fig. S1: Curie-constants $C$ (a) and Pauli-paramagnetic susceptibilities $\chi_0$ (b) as functions of tolerance factor $t$, calculated from the magnetic susceptibilities of the high-temperature phase; the triangle, round and square symbols represent $La_{4-x}Pr_xNi_3O_{10}$ ($x = 0, 1, 2, 3$), $Pr_{4-x}Nd_xNi_3O_{10}$ ($x = 0, 1, 2, 3$) and $Nd_{4-x}La_xNi_3O_{10}$ ($x = 0, 1, 2, 3$), respectively.

*Estimation of possible magnetic impurities due to Ni*

According to our X-ray data, we could not identify any impurity phases. To obtain an estimate of an upper bound of contributions to the magnetic susceptibility due to Ni-containing impurities, we refer here to the $\chi(T)$ data of $La_4Ni_3O_{10}$ where a significant contribution due to Ni impurities should be most visible. We performed various fits (see Figure S2), and obtain in the worst scenario a contribution to $C \approx 0.01$ emu K Oe$^{-1}$ mol$^{-1}$ at low temperatures, corresponding to $0.025 \approx$ spin ½ or $\approx 0.008$ spin-1 magnetic moments per formula unit. Considering the identical preparation conditions for all samples and assuming similar contamination levels, we can state that the values for $C$ ranging between $\approx 1.2$ and $7.7$ emu Oe$^{-1}$ mol$^{-1}$ are hardly affected by possible Ni impurities.

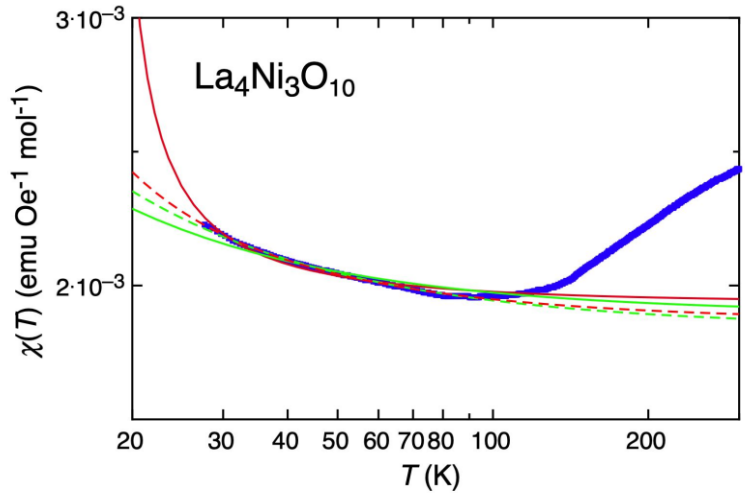

| $\chi(T) = C/(T+\Theta) + \chi_0$ | | | | |
|---|---|---|---|---|
| Fitting range 25K - 140 K | $C$ = 3.14e-3 | $\Delta C$ = 2.53-4 | | |
| | $\Theta$ = -17.6 | $\Delta\Theta$ = 0.93 | | |
| | $\chi_0$ = 1.94e-3 | $\Delta\chi_0$ = 3.66e-6 | | |
| Fitting range 25K - 100 K | $C$ = 7.65e-3 | $\Delta C$ = 2.33-4 | | |
| | $\Theta$ = -6.27 | $\Delta\Theta$ = 0.59 | | |
| | $\chi_0$ = 1.87e-3 | $\Delta\chi_0$ = 2.62e-6 | | |
| $\chi(T) = C/T + \chi_0$ | | | | |
| Fitting range 25K - 140 K | $C$ = 7.83e-3 | $\Delta C$ = 2.35-4 | | |
| | $\chi_0$ = 1.90e-3 | $\Delta\chi_0$ = 3.80e-6 | | |
| Fitting range 25K - 100 K | $C$ = 1.02e-2 | $\Delta C$ = 7.3e-5 | | |
| | $\chi_0$ = 1.84e-3 | $\Delta\chi_0$ = 1.39e-6 | | |

Fig. S2: Results from various Curie-Weiss like fits to the magnetic susceptibility of $La_4Ni_3O_{10}$ to estimate the contribution of magnetic impurities.

Table SI Crystallographic parameters obtained from the Rietveld refinements of $Ln_{4-x}Ln'_xNi_3O_{10}$ (*Ln* and *Ln'* = La, Pr and Nd; $x$ = 0, 1, 2, 3) at room temperature as well as their oxygen contents as determined from thermo-gravimetric measurements.

|  | a (Å) | b (Å) | c (Å) | β (°) | $R_p$ (%) | $R_{wp}$ (%) | $\chi^2$ | O content |
|---|---|---|---|---|---|---|---|---|
| $La_4Ni_3O_{10}$ | 5.403 | 5.454 | 14.24 | 100.70 | 4.89 | 9.08 | 12.6 | 10.03 |
| $La_3Pr_1Ni_3O_{10}$ | 5.396 | 5.458 | 14.17 | 100.71 | 2.69 | 4.27 | 3.01 | 10.10 |
| $La_2Pr_2Ni_3O_{10}$ | 5.382 | 5.457 | 14.12 | 100.71 | 3.19 | 5.27 | 4.71 | 10.01 |
| $La_1Pr_3Ni_3O_{10}$ | 5.382 | 5.466 | 14.07 | 100.78 | 2.56 | 4.15 | 5.14 | 10.00 |
| $Pr_4Ni_3O_{10}$ | 5.372 | 5.458 | 14.02 | 100.81 | 2.32 | 3.59 | 5.28 | 9.97 |
| $Pr_3Nd_1Ni_3O_{10}$ | 5.361 | 5.451 | 13.98 | 100.79 | 2.18 | 3.27 | 2.1 | 9.95 |
| $Pr_2Nd_2Ni_3O_{10}$ | 5.358 | 5.449 | 13.96 | 100.78 | 2.35 | 3.63 | 2.45 | 9.93 |
| $Pr_1Nd_3Ni_3O_{10}$ | 5.356 | 5.447 | 13.95 | 100.80 | 2.19 | 3.26 | 1.94 | 10.07 |
| $Nd_4Ni_3O_{10}$ | 5.351 | 5.441 | 13.93 | 100.79 | 2.52 | 3.85 | 2.67 | 9.93 |
| $La_1Nd_3Ni_3O_{10}$ | 5.363 | 5.459 | 14.08 | 100.73 | 5.33 | 9.14 | 9.66 | 9.98 |
| $La_2Nd_2Ni_3O_{10}$ | 5.375 | 5.463 | 14.14 | 100.67 | 5.67 | 10 | 13.8 | 9.99 |
| $La_3Nd_1Ni_3O_{10}$ | 5.401 | 5.466 | 14.19 | 100.68 | 4.14 | 6.94 | 8.01 | 9.95 |

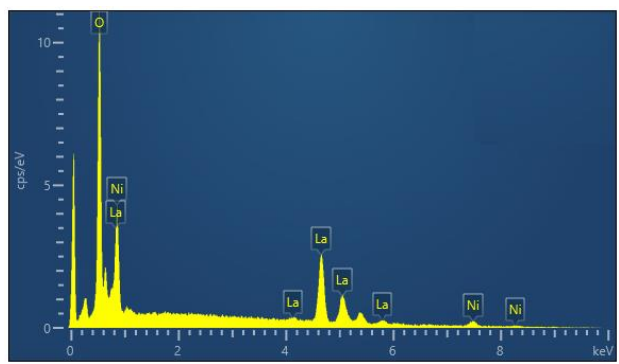

Fig. S3: EDX spectrum of $La_4Ni_3O_{10}$.

| Point-1 | | | Point-2 | | | Point-3 | | |
|---|---|---|---|---|---|---|---|---|
| Element | Weight% | Atomic % | Element | Weight% | Atomic% | Element | Weight% | Atomic% |
| La | 63.5 | 25 | La | 64.1 | 25.4 | La | 63.2 | 24.8 |
| Ni | 20.1 | 18.7 | Ni | 19.5 | 18.3 | Ni | 20.4 | 18.9 |
| O | 26.1 | 56.3 | O | 16.4 | 56.3 | O | 16.5 | 56.2 |
| Total | 100 | | Total | 100 | | Total | 100 | |
| Point-4 | | | Point-5 | | | | | |
| Element | Weight% | Atomic% | Element | Weight% | Atomic% | **Average rate of La/Ni: 4.01(6)/3** | | |
| La | 63.5 | 25 | La | 63.3 | 24.9 | | | |
| Ni | 20.1 | 18.7 | Ni | 20.3 | 18.9 | | | |
| O | 16.4 | 56.3 | O | 16.5 | 56.2 | | | |
| Total | 100 | | Total | 100 | | | | |

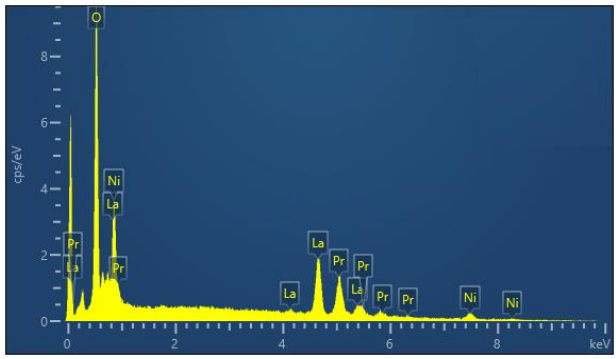

Fig. S4: EDX spectrum of $La_3Pr_1Ni_3O_{10}$.

| Point-1 | | | Point-2 | | | Point-3 | | |
|---|---|---|---|---|---|---|---|---|
| Element | Weight% | Atomic % | Element | Weight% | Atomic% | Element | Weight% | Atomic% |
| Pr | 16.1 | 6.2 | Pr | 17.2 | 6.7 | Pr | 16.6 | 6.4 |
| La | 46.4 | 18.2 | La | 47 | 18.6 | La | 46.4 | 18.3 |
| Ni | 21 | 19.5 | Ni | 19.5 | 18.3 | Ni | 20.6 | 19.2 |
| O | 16.1 | 56.1 | O | 16.4 | 56.3 | O | 16.4 | 56.2 |
| Total | 100 | | Total | 100 | | Total | 100 | |
| Point-4 | | | Point-5 | | | | | |
| Element | Weight% | Atomic% | Element | Weight% | Atomic% | **Average rate of Pr/ La/Ni:** | | |
| Pr | 17 | 6.6 | Pr | 16.8 | 6.5 | **1.01(4)/2.84(9)/3** | | |
| La | 45.5 | 17.8 | La | 45.8 | 17.9 | | | |
| Ni | 21.1 | 19.5 | Ni | 21.1 | 19.5 | | | |
| O | 16.5 | 5.1 | O | 16.5 | 56.1 | | | |
| Total | 100 | | Total | 100 | | | | |

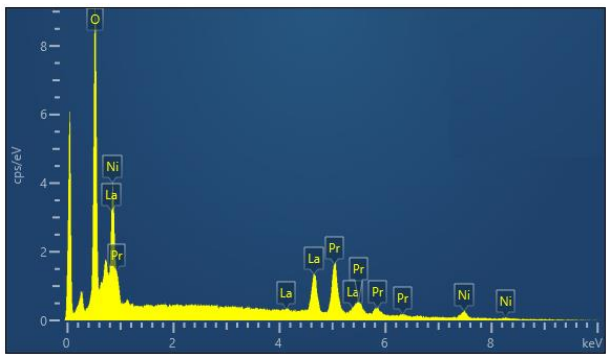

Fig. S5: EDX spectrum of $La_2Pr_2Ni_3O_{10}$.

| Point-1 | | | Point-2 | | | Point-3 | | |
|---|---|---|---|---|---|---|---|---|
| Element | Weight% | Atomic % | Element | Weight% | Atomic% | Element | Weight% | Atomic% |
| Pr | 33.9 | 13.3 | Pr | 33.7 | 13.1 | Pr | 34.1 | 13.3 |
| La | 29.8 | 11.8 | La | 29.4 | 11.6 | La | 29.8 | 11.8 |

| Ni | 19.9 | 18.6 | Ni | 20.4 | 19.1 | Ni | 29.8 | 18.5 |
|---|---|---|---|---|---|---|---|---|
| O | 16.4 | 56.3 | O | 16.4 | 56.2 | O | 16.3 | 56.3 |
| Total | 100 | | Total | 100 | | Total | 100 | |
| Point-4 | | | Point-5 | | | | | |
| Element | Weight% | Atomic% | Element | Weight% | Atomic% | **Average rate of Pr/ La/Ni:** | | |
| Pr | 33.9 | 13.1 | Pr | 34.2 | 13.4 | | | |
| La | 28.9 | 11.4 | La | 29.9 | 11.9 | **2.11(6)/1.87(6)/3** | | |
| Ni | 20.8 | 19.3 | Ni | 19.6 | 18.4 | | | |
| O | 16.4 | 26.1 | O | 16.3 | 56.3 | | | |
| Total | 100 | | Total | 100 | | | | |

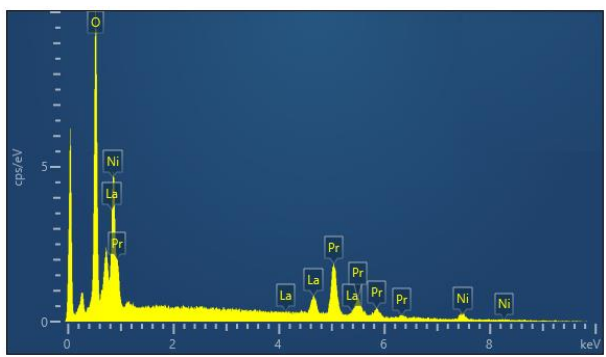

Fig. S6: EDX spectrum of $La_1Pr_3Ni_3O_{10}$.

| Point-1 | | | Point-2 | | | Point-3 | | |
|---|---|---|---|---|---|---|---|---|
| Element | Weight% | Atomic % | Element | Weight% | Atomic% | Element | Weight% | Atomic% |
| Pr | 48.4 | 18.8 | Pr | 49.1 | 19.2 | Pr | 48.1 | 18.7 |
| La | 14.6 | 5.8 | La | 14.3 | 5.7 | La | 14.7 | 5.8 |
| Ni | 20.6 | 19.3 | Ni | 20.1 | 18.9 | Ni | 20.8 | 19.4 |
| O | 16.4 | 56.1 | O | 16.3 | 56.2 | O | 16.4 | 56.1 |
| Total | 100 | | Total | 100 | | Total | 100 | |
| Point-4 | | | Point-5 | | | | | |
| Element | Weight% | Atomic% | Element | Weight% | Atomic% | | | |
| Pr | 48.7 | 19 | Pr | 47.7 | 18.6 | **Average rate of Pr/ La/Ni:** | | |
| La | 14.5 | 5.8 | La | 15.4 | 6.1 | **2.95(6)/0.91(2)/3** | | |
| Ni | 20.4 | 19.1 | Ni | 20.5 | 19.2 | | | |
| O | 16.4 | 56.2 | O | 16.4 | 56.2 | | | |
| Total | 100 | | Total | 100 | | | | |

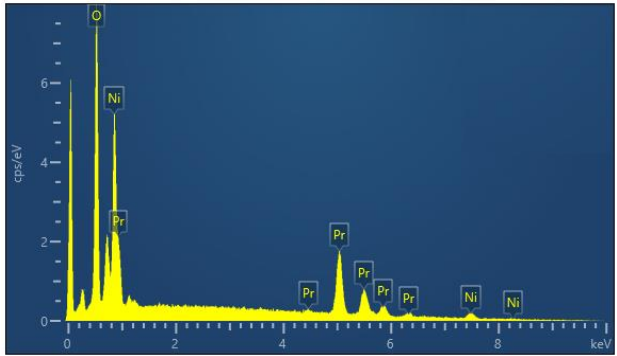

Fig. S7: EDX spectrum of $Pr_4Ni_3O_{10}$.

| Point-1 | | | Point-2 | | | Point-3 | | |
|---|---|---|---|---|---|---|---|---|
| Element | Weight% | Atomic % | Element | Weight% | Atomic% | Element | Weight% | Atomic% |
| Pr | 63.5 | 24.8 | Pr | 63.8 | 25 | Pr | 64 | 25.1 |
| Ni | 20.2 | 19 | Ni | 20 | 18.8 | Ni | 19.7 | 18.6 |
| O | 16.3 | 56.2 | O | 16.3 | 56.2 | O | 16.3 | 56.3 |
| Total | 100 | | Total | 100 | | Total | 100 | |
| Point-4 | | | Point-5 | | | | | |
| Element | Weight% | Atomic% | Element | Weight% | Atomic% | **Average rate of Pr/Ni: 4.01(5)/3** | | |
| Pr | 64.2 | 25.2 | Pr | 63.9 | 25.1 | | | |
| Ni | 19.6 | 18.5 | Ni | 19.8 | 18.7 | | | |
| O | 16.3 | 56.3 | O | 16.3 | 56.3 | | | |
| Total | 100 | | Total | 100 | | | | |

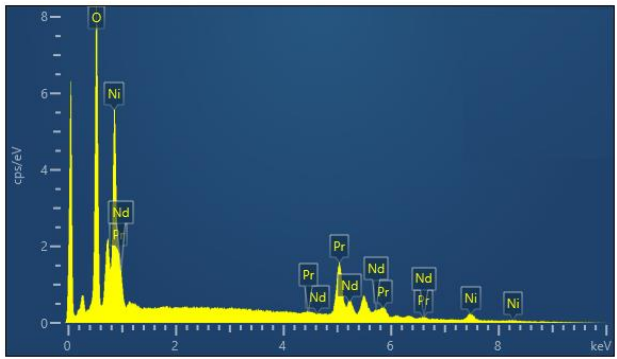

Fig. S8: EDX spectrum of $Pr_3Nd_1Ni_3O_{10}$.

| Point-1 | | | Point-2 | | | Point-3 | | |
|---|---|---|---|---|---|---|---|---|
| Element | Weight% | Atomic % | Element | Weight% | Atomic% | Element | Weight% | Atomic% |
| Pr | 48 | 18.9 | Pr | 48.1 | 18.9 | Pr | 49.1 | 19.4 |
| Nd | 15.7 | 6 | Nd | 15.4 | 6 | Nd | 15.3 | 5.9 |
| Ni | 20 | 18.9 | Ni | 20 | 18.8 | Ni | 19.4 | 18.4 |
| O | 16.2 | 56.2 | O | 16.2 | 56.2 | O | 16.2 | 56.3 |
| Total | 100 | | Total | 100 | | Total | 100 | |

| Point-4 | | | Point-5 | | | |
|---|---|---|---|---|---|---|
| Element | Weight% | Atomic% | Element | Weight% | Atomic% | |
| Pr | 47.2 | 18.5 | Pr | 47.1 | 18.5 | **Average rate of Pr/ Nd/Ni:** |
| Nd | 16.2 | 6.2 | Nd | 16.3 | 6.2 | **3.00(8)/0.96(1)/3** |
| Ni | 20.4 | 19.2 | Ni | 20.1 | 19.1 | |
| O | 16.3 | 56.2 | O | 16.3 | 56.2 | |
| Total | 100 | | Total | 100 | | |

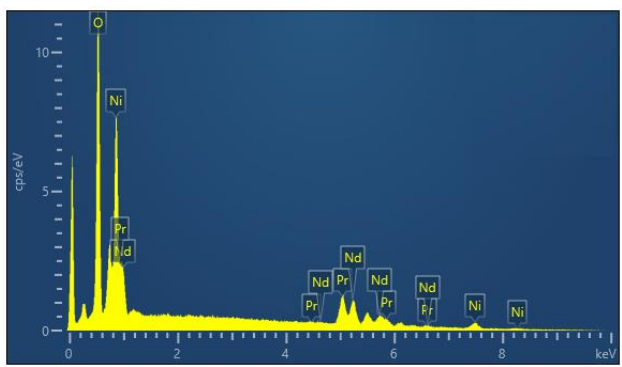

Fig. S9: EDX spectrum of $Pr_2Nd_2Ni_3O_{10}$.

| Point-1 | | | Point-2 | | | Point-3 | | |
|---|---|---|---|---|---|---|---|---|
| Element | Weight% | Atomic % | Element | Weight% | Atomic% | Element | Weight% | Atomic% |
| Pr | 31.8 | 12.5 | Pr | 32 | 12.7 | Pr | 32.3 | 12.3 |
| Nd | 32 | 12.3 | Nd | 32.1 | 12.4 | Nd | 31.9 | 12.8 |
| Ni | 20 | 18.9 | Ni | 19.7 | 18.7 | Ni | 19.6 | 18.6 |
| O | 16.2 | 56.2 | O | 16.2 | 56.3 | O | 16.2 | 56.3 |
| Total | 100 | | Total | 100 | | Total | 100 | |

| Point-4 | | | Point-5 | | | |
|---|---|---|---|---|---|---|
| Element | Weight% | Atomic% | Element | Weight% | Atomic% | |
| Pr | 31.1 | 12.2 | Pr | 31.6 | 12.4 | **Average rate of Pr/ Nd/Ni:** |
| Nd | 32.2 | 12.4 | Nd | 31.7 | 12.2 | **1.97(5)/1.97(5)/3** |
| Ni | 20.4 | 19.3 | Ni | 20.5 | 19.3 | |
| O | 16.2 | 56.1 | O | 16.2 | 56.1 | |
| Total | 100 | | Total | 100 | | |

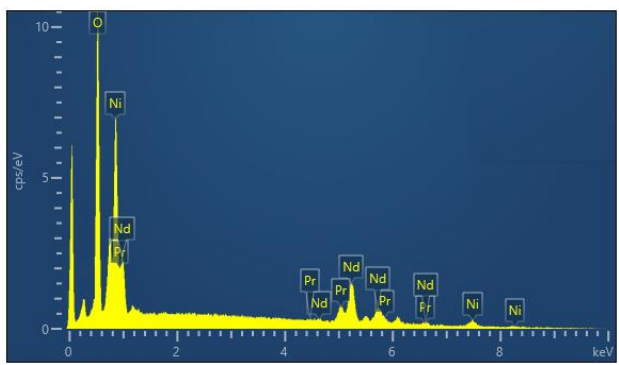

Fig. S10: EDX spectrum of $Pr_1Nd_3Ni_3O_{10}$.

| Point-1 | | | Point-2 | | | Point-3 | | |
|---|---|---|---|---|---|---|---|---|
| Element | Weight% | Atomic % | Element | Weight% | Atomic% | Element | Weight% | Atomic% |
| Pr | 15.8 | 6.4 | Pr | 16.7 | 6.2 | Pr | 16.7 | 6.6 |
| Nd | 47.8 | 19 | Nd | 48.1 | 19.2 | Nd | 47.6 | 18.6 |
| Ni | 20.3 | 18.4 | Ni | 19.2 | 18.4 | Ni | 19.6 | 18.4 |
| O | 16.2 | 56.2 | O | 16.1 | 56.2 | O | 16.1 | 56.3 |
| Total | 100 | | Total | 100 | | Total | 100 | |
| Point-4 | | | Point-5 | | | | | |
| Element | Weight% | Atomic% | Element | Weight% | Atomic% | **Average rate of Pr/ Nd/Ni:** | | |
| Pr | 15.7 | 6.2 | Pr | 15.4 | 6.1 | **1.02(3)/3.08(4)/3** | | |
| Nd | 48.2 | 18.9 | Nd | 48.1 | 19.2 | | | |
| Ni | 20 | 18.6 | Ni | 20.3 | 18.5 | | | |
| O | 16.1 | 56.2 | O | 16.2 | 56.2 | | | |
| Total | 100 | | Total | 100 | | | | |

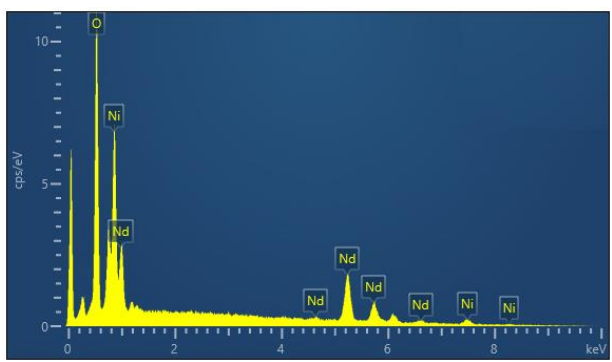

Fig. S11: EDX spectrum of $Nd_4Ni_3O_{10}$.

| Point-1 | | | Point-2 | | | Point-3 | | |
|---|---|---|---|---|---|---|---|---|
| Element | Weight% | Atomic % | Element | Weight% | Atomic% | Element | Weight% | Atomic% |
| Nd | 64.3 | 25.0 | Nd | 64.8 | 25.3 | Nd | 64.3 | 25.0 |
| Ni | 19.7 | 18.8 | Ni | 19.2 | 18.4 | Ni | 19.7 | 18.8 |
| O | 16.1 | 56.2 | O | 16.0 | 56.3 | O | 16.1 | 56.2 |
| Total | 100 | | Total | 100 | | Total | 100 | |
| Point-4 | | | Point-5 | | | | | |
| Element | Weight% | Atomic% | Element | Weight% | Atomic% | **Average rate of Nd/Ni:** | | |
| Nd | 64.6 | 25.2 | Nd | 64.4 | 25.0 | **4.04(6)/3** | | |
| Ni | 19.4 | 18.5 | Ni | 19.6 | 18.7 | | | |
| O | 16.0 | 56.3 | O | 16.0 | 56.3 | | | |
| Total | 100 | | Total | 100 | | | | |

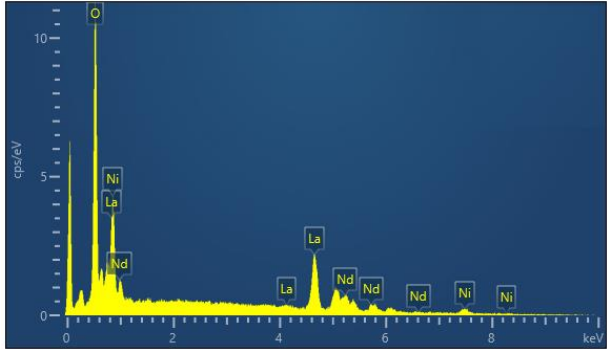

Fig. S12: EDX spectrum of $La_3Nd_1Ni_3O_{10}$.

| Point-1 | | | Point-2 | | | Point-3 | | |
|---|---|---|---|---|---|---|---|---|
| Element | Weight% | Atomic % | Element | Weight% | Atomic% | Element | Weight% | Atomic% |
| La | 47.0 | 18.6 | La | 46.1 | 17.9 | La | 45.5 | 18.9 |
| Nd | 16.9 | 6.5 | Nd | 16.9 | 6.9 | Nd | 16.1 | 16.7 |
| Ni | 19.8 | 18.7 | Ni | 20.5 | 18.9 | Ni | 22.0 | 18.1 |
| O | 16.3 | 56.2 | O | 16.5 | 56.3 | O | 16.4 | 56.3 |
| Total | 100 | | Total | 100 | | Total | 100 | |
| Point-4 | | | Point-5 | | | | | |
| Element | Weight% | Atomic% | Element | Weight% | Atomic% | **Average rate of La/Nd/Ni:** | | |
| La | 45.1 | 18.4 | La | 46.0 | 18.1 | **2.9(1)/1.04(5)/3** | | |
| Nd | 18.5 | 16.3 | Nd | 16.5 | 6.3 | | | |
| Ni | 20.0 | 19.1 | Ni | 21.2 | 19.4 | | | |
| O | 16.4 | 56.2 | O | 16.3 | 56.2 | | | |
| Total | 100 | | Total | 100 | | | | |

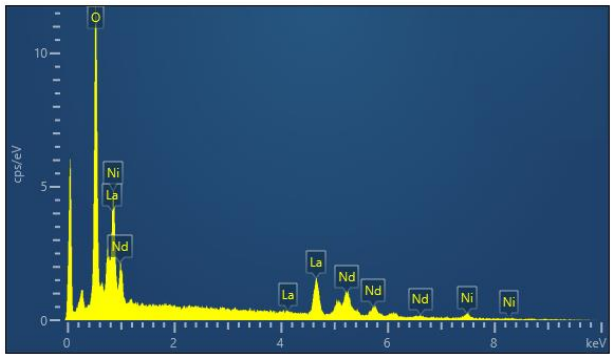

Fig. S13: EDX spectrum of $La_2Nd_2Ni_3O_{10}$.

| Point-1 | | | Point-2 | | | Point-3 | | |
|---|---|---|---|---|---|---|---|---|
| Element | Weight% | Atomic % | Element | Weight% | Atomic% | Element | Weight% | Atomic% |
| La | 30.5 | 12.2 | La | 30.0 | 11.9 | La | 31.1 | 12.6 |
| Nd | 33.0 | 12.6 | Nd | 33.0 | 12.6 | Nd | 33.4 | 13.0 |
| Ni | 20.1 | 19.0 | Ni | 20.7 | 19.3 | Ni | 19.2 | 18.2 |

| O | 16.4 | 56.3 | O | 16.3 | 56.2 | O | 16.3 | 56.2 |
| --- | --- | --- | --- | --- | --- | --- | --- | --- |
| Total | 100 | | Total | 100 | | Total | 100 | |
| Point-4 | | | Point-5 | | | | | |
| Element | Weight% | Atomic% | Element | Weight% | Atomic% | **Average rate of La/Nd/Ni: 1.95(6)/2.01(5)/3** | | |
| La | 31.2 | 12.4 | La | 30.4 | 12.1 | | | |
| Nd | 32.3 | 12.5 | Nd | 32.7 | 12.5 | | | |
| Ni | 20.1 | 18.8 | Ni | 20.5 | 19.1 | | | |
| O | 16.4 | 56.3 | O | 16.4 | 56.3 | | | |
| Total | 100 | | Total | 100 | | | | |

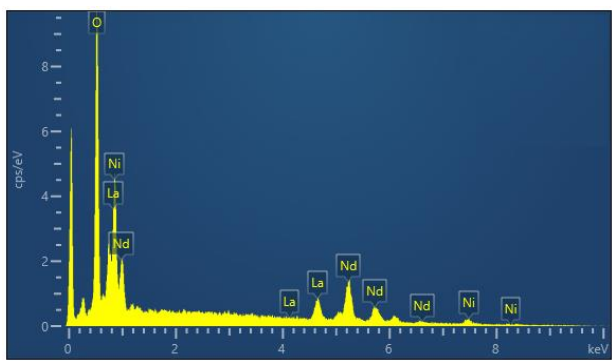

Fig. S14: EDX spectrum of $La_1Nd_3Ni_3O_{10}$.

| Point-1 | | | Point-2 | | | Point-3 | | |
| --- | --- | --- | --- | --- | --- | --- | --- | --- |
| Element | Weight% | Atomic % | Element | Weight% | Atomic% | Element | Weight% | Atomic% |
| La | 15.1 | 6.1 | La | 14.9 | 6.0 | La | 15.1 | 6.1 |
| Nd | 49.5 | 19.2 | Nd | 49.9 | 19.4 | Nd | 19.2 | 19.1 |
| Ni | 19.3 | 18.4 | Ni | 19.1 | 18.2 | Ni | 19.6 | 18.7 |
| O | 16.1 | 56.3 | O | 16.1 | 56.4 | O | 16.1 | 56.2 |
| Total | 100 | | Total | 100 | | Total | 100 | |
| Point-4 | | | Point-5 | | | | | |
| Element | Weight% | Atomic% | Element | Weight% | Atomic% | **Average rate of La/Nd/Ni: 0.98(3)/3.0(1)/3** | | |
| La | 15.1 | 5.9 | La | 16.2 | 6.5 | | | |
| Nd | 49.5 | 18.4 | Nd | 48.1 | 18.6 | | | |
| Ni | 19.3 | 19.6 | Ni | 19.5 | 18.5 | | | |
| O | 16.1 | 56.1 | O | 16.2 | 56.3 | | | |
| Total | 100 | | Total | 100 | | | | |